\documentclass[twoside]{hs04}
\def\runauthor{O.V.Utyuzh, G.Wilk, Z.W\l odarczyk}
\def\shorttitle{BEC in the Quantum Clan approach}
\begin{document}

\title{BOSE-EINSTEIN CORRELATIONS\\
IN THE QUANTUM CLAN APPROACH }

\author{O.V.UTYUZH and G.WILK}
\address{The Andrzej So\l tan Institute for Nuclear Studies, Zd-PVIII\\
Ho\.za 69, 00-681 Warszawa, Poland\\
E-mails: utyuzh@fuw.edu.pl and wilk@fuw.edu.pl}
\author{Z.W\L ODARCZYK}
\address{Institute of Physics, \'Swi\c{e}tokrzyska Academy\\
\'Swi\c{e}tokrzyska 15, 25-406 Kielce, Poland\\
E-mail: wlod@pu.kielce.pl}

\maketitle

\abstracts{We propose novel numerical method of {\it modelling}
Bose-Einstein correlations (BEC) observed among identical (bosonic)
particles produced in multiparticle production reactions. We argue
that the most natural approach is to work directly in the momentum
space in which the Bose statistics of secondaries reveals itself in
their tendency to bunch in a specific way in the available phase
space. Because such procedure is essentially identical to the clan
model of multiparticle distributions proposed some time ago,
therefore we call it the {\it Quantum Clan Model}. }


The phenomenon of Bose-Einstein correlations (BEC) is so widely know
that we shall not introduce it here again referring instead to
\cite{BEC}. We shall instead address problem of the {\it proper
numerical modelling of BEC} understood as approach which accounts
from the very beginning for the quantum statistical bosonic character
of identical secondaries produced in hadronization process. To our
knowledge this problem was so far considered only in \cite{OMT}(cf.,
however, \cite{ZAJC}). All other approaches claiming to model BEC
numerically \cite{modBEC} use as their starting point the outcomes of
existing Monte-Carlo event generators (MCG) describing multiparticle
production process \cite{GEN} and {\it modify} them in a suitable way
to fit the BEC data. These modifications are called {\it
afterburners}. They inevitably lead to such unwanted features as
violation of energy-momentum conservation or to changes in the
original multiparticle spectra.

In \cite{UWW} we have proposed afterburner free from such unwanted
effects. It was based on different concept of introducing quantum
mechanical (QM) effects in the otherwise purely probabilistic
distributions from those proposed in \cite{QUANT}. Namely, each MCG
provides us usually with a given number of particles, each one
endowed with one of $(+/-/0)$ charge and with well defined
spatio-temporal position and energy-momentum. On the other hand
experiment provides us information on only the first and last
characteristics. The spatio-temporal information is not available
directly (in fact, the universal hope expressed in \cite{BEC,modBEC}
is it can be deduced from the previous two via the measured BEC). Our
reasoning was as follows: $(i)$ BEC phenomenon is of the QM origin,
therefore one has to introduce in the otherwise purely classical
distributions provided by MCG a new element mimicking QM
uncertainties; $(ii)$ this cannot be done with energy-momenta because
they are measured and therefore fixed; $(iii)$ the next candidate,
i.e., spatio-temporal characteristics, can be changed but this was
already done in \cite{QUANT,modBEC}; $(iv)$ one is thus left with
charges and in \cite{UWW} we have simply assigned (on event-by-event
basis) new charges to the particles selected by MCG conserving,
however, the original multiplicity of $(+/-/0)$. This has been done
in such way as to make particles of the same charge to be located
maximally near to each other in the phase space by exploring natural
fluctuations of spatio-temporal and energy-momentum characteristic
resulting from MCG. In this way automatically conserve all
energy-momenta and do not change multiparticle distributions and do
it already on the {\it level of each event} provided by MCG .
However, the new assignment of charges introduces a profound change
in the structure of the original MCG. Generally speaking (cf.
\cite{UWW} for details) it is equivalent to introduction of bunching
of particles of the same charge.

This observation will be the cornerstone of our new proposition. Let
us remind that idea of bunching of particles as quantum statistical
(QS) effect is not new \cite{QS}. It was used in connection with BEC
for the first time in \cite{GN} and then was a cornerstone of the
{\it clan model} of multiparticle distributions $P(n)$ leading in
natural way to their negative binomial (NB) form observed in
experiment \cite{NBD}. It was introduced in the realm of BEC in
\cite{BSWW} and \cite{OMT,ZAJC}. Because our motivation comes
basically from \cite{OMT} let us outline shortly its basic points. It
deals with the problem of how to distribute in a least biased way a
given number of bosonic secondaries, $\langle n\rangle =\langle
n^{(+)}\rangle + \langle n^{(-)}\rangle + \langle n^{(0)}\rangle$,
$\langle n^{(+)}\rangle =\langle n^{(-)}\rangle =\langle
n^{(0)}\rangle$. Using information theory approach (cf., \cite{IT})
their rapidity distribution was obtained in form of grand partition
function with temperature $T$ and chemical potential $\mu$. In
addition, the rapidity space was divided into {\it cells} of size
$\delta y$ (fitted parameter) each. It turned out that whereas the
very fact of existence of such cells was enough to obtain reasonably
good multiparticle distributions, $P(n)$, (actually, in the NB-like
form), their size, $\delta y$, was crucial for obtaining the
characteristic form of the $2-$body BEC function $C_2(Q=|p_i-p_j|)$
(peaked and greater than unity at $Q=0$ and then decreasing in a
characteristic way towards $C_2=1$ for large values of $Q$) out of
which one usually deduces the spatio-temporal characteristics of the
hadronization source \cite{BEC} (see \cite{OMT} for more details).
The outcome was obvious: to get $C_2$ peaked and greater than unity
at $Q=0$ and then decreasing in a characteristic way towards $C_2=1$
for large values of $Q$ one must have particles located in cells in
phase space which are of nonzero size\footnote{It means then that
from $C_2$ one gets not the size of the hadronizing source but only
size of the emitting cell, in \cite{OMT} $R\sim 1/\delta y$, cf.
\cite{Z}. In the quantum field theoretical formulation of BEC this
directly corresponds to the necessity of replacing delta functions in
commutator relations by a well defined peaked functions introducing
in this way same dimensional scale to be obtained from fits to data
\cite{Kozlov}. This fact was known even before but without any
phenomenological consequences \cite{Zal}.}.

\begin{figure}[ht]
\hspace{1.15cm}
  \begin{minipage}[ht]{104mm}
    \centerline{
        \epsfig{file=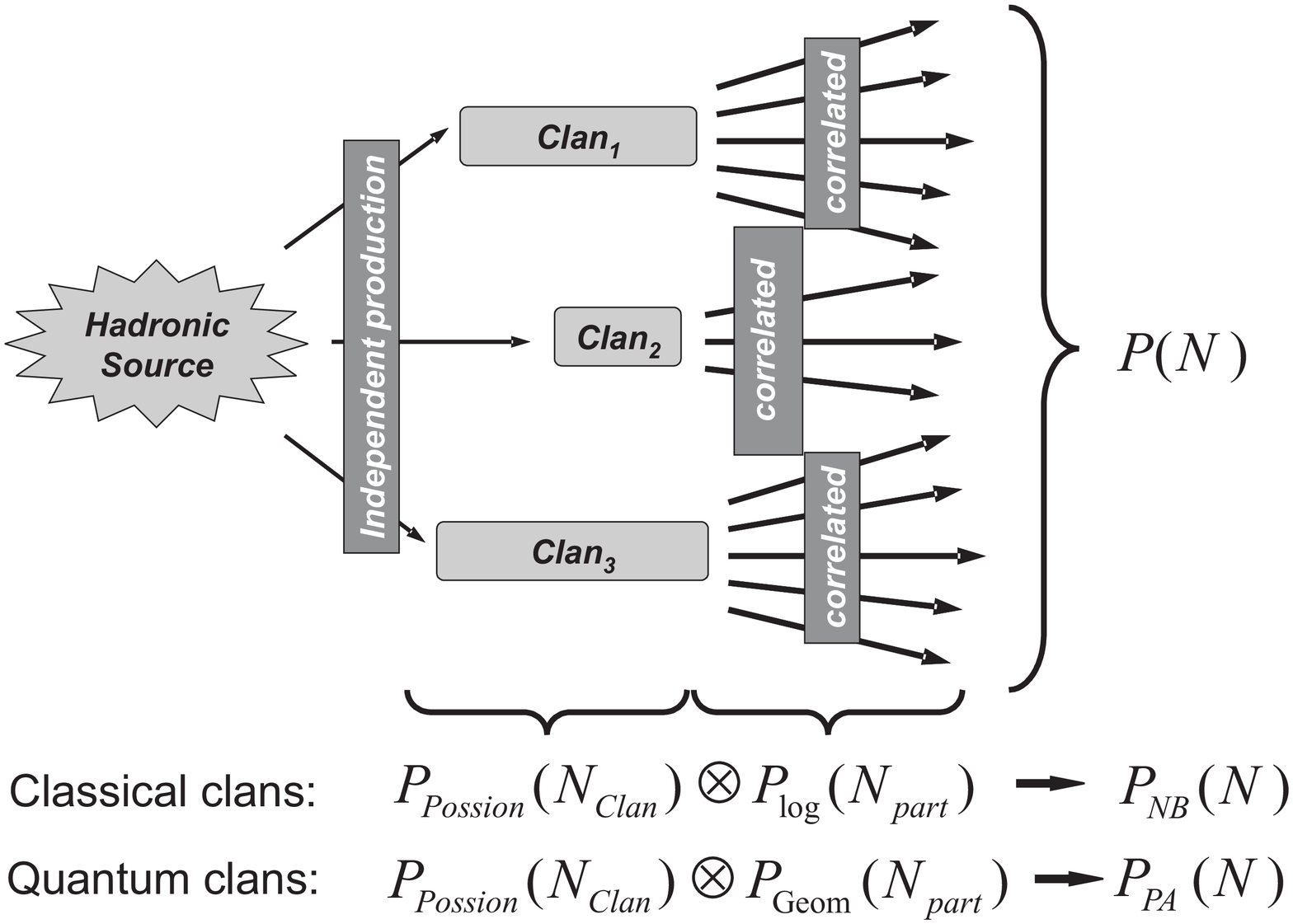, width=90mm}
     }
  \end{minipage}
  \vspace{0.5cm}
  \hfill
  \begin{minipage}[ht]{57mm}
    \centerline{
        \epsfig{file=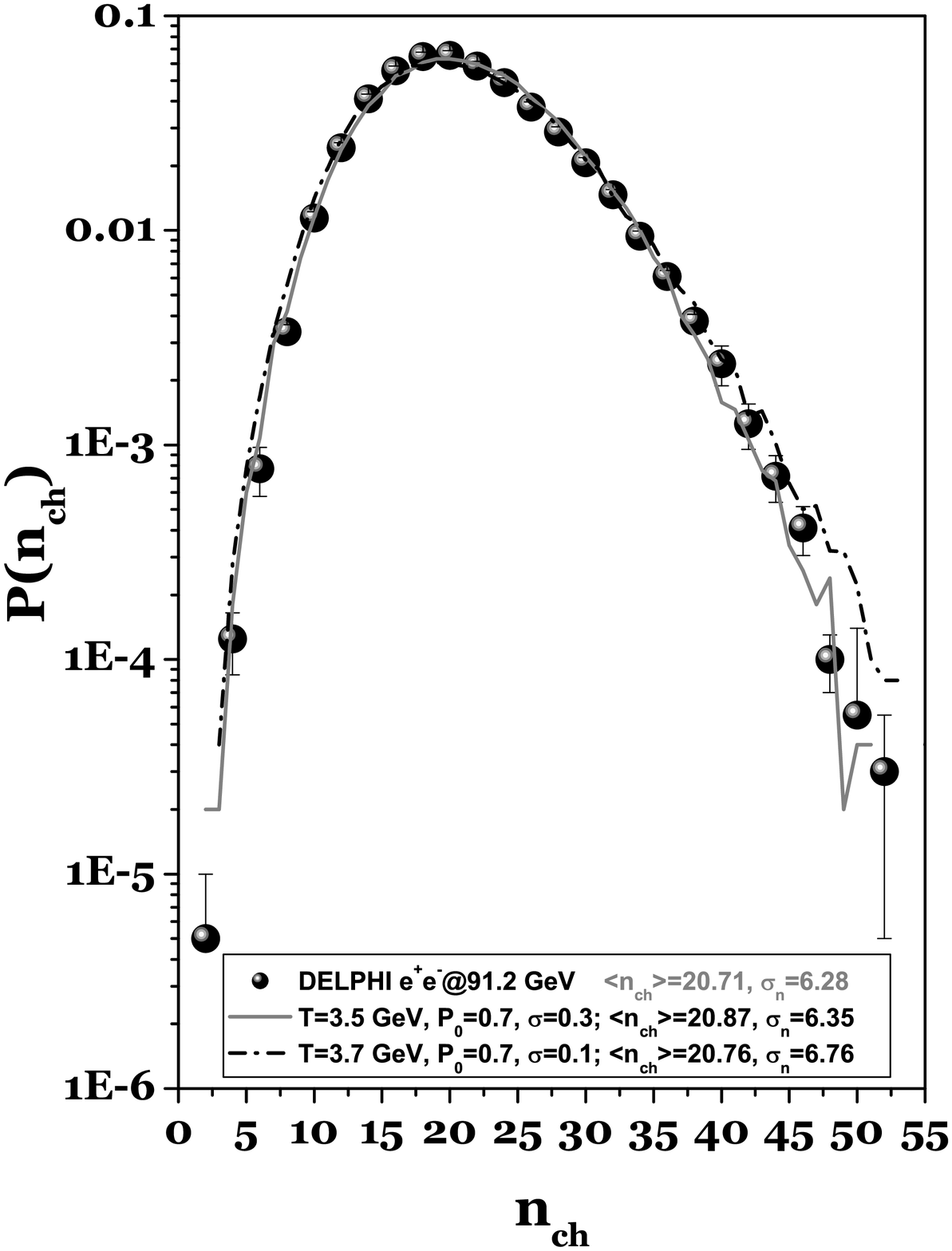, width=55mm}
     }
  \end{minipage}
\hfill
  \begin{minipage}[ht]{57mm}
    \centerline{
       \epsfig{file=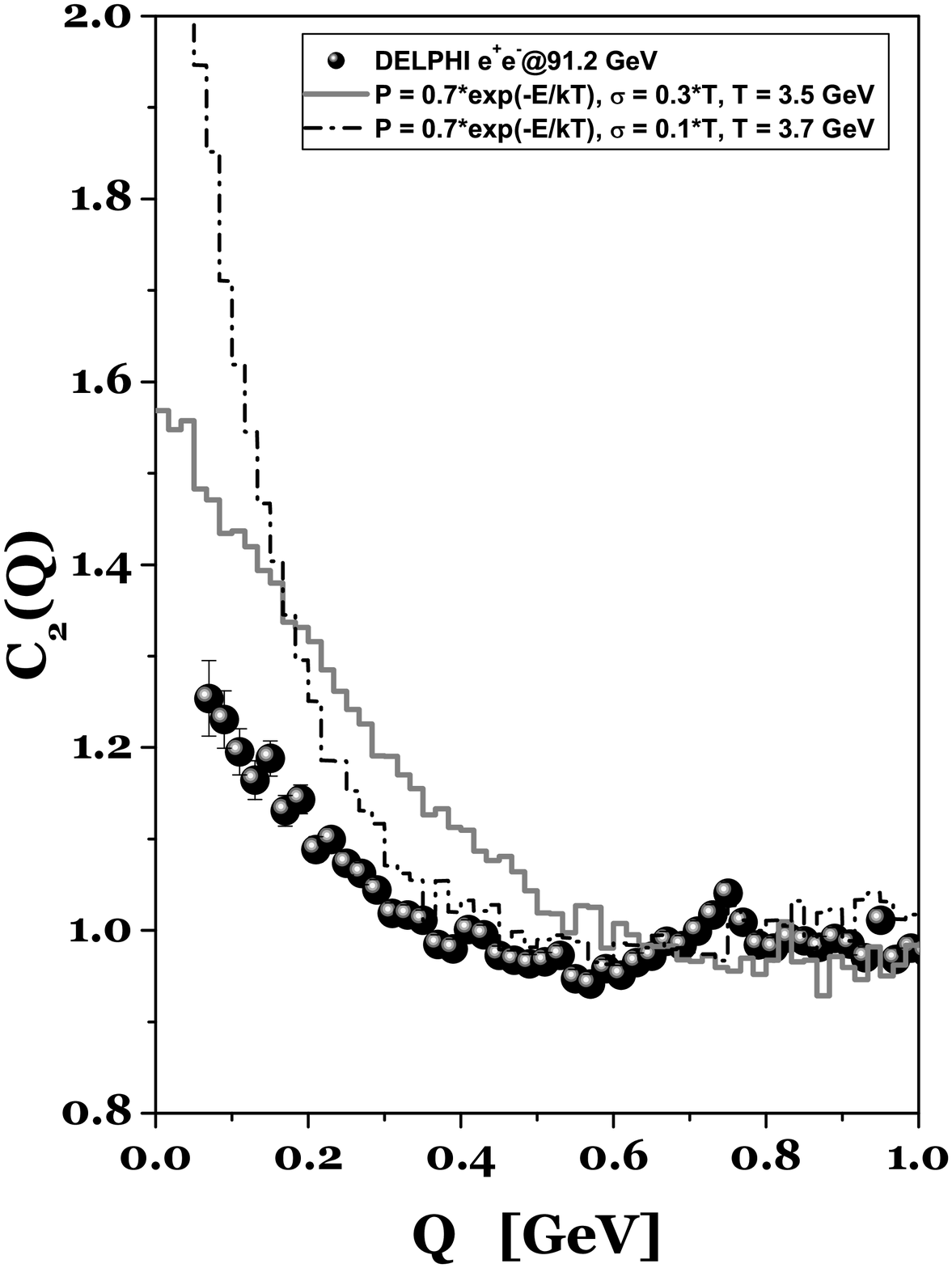, width=55mm}
     }
  \end{minipage}
  \caption{
\footnotesize {Upper panel: the proposed algorithm is similar to the
classical clan model proposed in \protect\cite{NBD} but its clans
contain particles of the same charge and (almost) the same energies
and are distributed according to geometrical distribution what
results in overall P\`olya-Aeppli distribution, $P_{PA}(n)$
\protect\cite{PA}, insted of NB one, $P_{NB}(n)$. Lower panels
contain examples of our results (data from \protect\cite{Data} were
used for comparison). Left panel: fit to charge multiplicity
distribution. Right panel: results for $C_2(Q=|p_1-p_2|)$ correlation
function (one dimensional phase space was used here only). Two
different sets of parameters have been used. Notice that whereas they
lead to essentially similar $P(n)$ the resulting $C_2(Q)$ are
drastically different.}}
  \label{fig:Fig1}
\end{figure}

To illustrate our proposition assume that mass $M$ hadronizes into
$N=\langle n\rangle$ bosonic particles (we take them as pions of mass
$m$) with equal numbers of $(+/-/0)$ charges and with limited
transverse momenta $p_T$. Suppose that their multiplicity
distribution $P(n)$ follows a NB-like form (i.e., it is broader than
Poissonian) and that their two-particle correlation function of
identical particles, $C_2(Q)$, has the specific BEC form mentioned
above. To model such process accounting from the very beginning, for
the bosonic character of produced particles we propose the following
steps (illustrated by comparison to some selected LEP $e^+e^-$ data
\cite{Data}, cf., Fig. \ref{fig:Fig1}):

{\bf (1)} Using some (assumed) function $f(E)$ select a particle of
energy $E^{(1)}_1$ and charge $Q^{(1)}$. The actual form of $f(E)$
should reflect somehow our {\it a priori} knowledge of the particular
collision process under consideration. In what follows we shall
assume that $f(E) = \exp\left( -E/T\right)$, with $T$ being parameter
(playing in our example the role of "temperature").

{\bf (2)} Treat this particle as seed of the first {\it elementary
emitting cell} (EEC) and add to it, until the first failure, other
particles of the same charge $Q^{(1)}$ selected according to
distribution $P(E)=P_0\cdot f(E)$, where $P_0$ is another parameter
(playing the role of "chemical potential" $\mu = T\cdot \ln P_0$). This
assures that the number of particles in this EEC, $k_1$, will follow
geometrical (or Bose-Einstein) distribution and accounts therefore
for their bosonic character. As result $C_2(Q)>1$ but only {\it at
one point}, namely for $Q=0$.

{\bf (3)} To get the experimentally observed width of $C_2(Q)$ one
has to allow that particles in each EEC can have (slightly) different
energies from energy of the particle being its seed. To do it allow
that each additional particle selected in point $(2)$ above have
energy $E^{(1)}_i$ selected from some distribution function peaked at
$E_1^{(1)}$, $G\left( E^{(1)}_1 - E^{(1)}_i\right)$, where the width
of this distribution, $\sigma$, is another free parameter.

{\bf (4)} Repeat points $(1)$ - $(3)$ as long as there is enough
energy left. Correct in every event for every energy-momentum
nonconservation caused by the selection procedure adopted and assure
that $N^{(+)}=N^{(-)}$.

As result in each event we get a number of EEC with particles of the
same charge and (almost) the same energy, i.e., picture closely
resembling classical {\it clans} of \cite{NBD} (with no effects of
statistics imposed, see Fig. \ref{fig:Fig1}). Clans are distributed
in the same way as the particles forming seeds for EEC, i.e.,
according to Poisson distribution. On the other hand, whereas in
\cite{NBD} particles in each clan were assumed to follow {\it
logarithmic} distribution, in our QM clans, or EEC, they are by
definition distributed according to geometrical distribution. As a
result the overall distribution of particles in our Quantum Clan
Model case will be of the so called P\`olya-Aeppli type \cite{PA}.
The first preliminary results presented in Fig. \ref{fig:Fig1} are
quite encouraging (especially when one remembers that so far effects
of resonances and all kind of final state interactions to which $C_2$
is sensitive were neglected here). It remains now to be checked what
two-body BEC functions for other components of the momentum
differences and how they depend on the EEC parameters: $T$, $P_0$ and
$\sigma$. So far the main outcome is that BEC are due to EEC's only
and therefore provide us mainly with their characteristics. This
should clear at least some of many apparently "strange" results
obtained from BEC recently.

\section*{Acknowledgements} Partial support of the Polish State
Committee for Scientific Research (KBN) (grant 2P03B04123 (ZW) and
grants 621/E-78/SPUB/CERN/P-03/DZ4/99 and 3P03B05724 (GW)) is
acknowledged.

\end{document}